\documentclass[a4paper,12pt]{article}
\usepackage[pctex32]{graphics}
\usepackage{amssymb,amsmath}
\usepackage{amsmath,amssymb}
\usepackage{latexsym}
\usepackage{epsfig}
\usepackage[english]{babel}

\newcommand{\be}{\begin{equation}}
\newcommand{\ee}{\end{equation}}
\newcommand{\ba}{\begin{eqnarray}}
\newcommand{\ea}{\end{eqnarray}}

\begin{document}

\begin{titlepage}

\vspace{5mm}

\begin{center}

{\Large \bf Propagating degrees of freedom \\ in f(R) gravity}

\vskip .6cm

\centerline{\large
 Yun Soo Myung$^{a}$}

\vskip .6cm

{Institute of Basic Science and Department of Computer Simulation,
\\Inje University, Gimhae 621-749, Korea \\}

\end{center}

\begin{center}
\underline{Abstract}
\end{center}
We have computed the number of polarization modes of gravitational
waves propagating in the Minkowski background in $f(R)$ gravity.
This is three of two from transverse-traceless tensor modes and one
from a massive trace mode, which confirms the results found in the
literature. There is no massless breathing mode and the massive
trace mode corresponds to the Ricci scalar. A newly defined metric
tensor in $f(R)$ gravity satisfies the transverse-traceless (TT)
condition as well as the TT wave equation.

 \vskip .6cm


\vskip 0.8cm

\vspace{15pt} \baselineskip=18pt

\noindent $^a$ysmyung@inje.ac.kr \\

\thispagestyle{empty}
\end{titlepage}

\newpage
%%%%%%%%%%%%%%%%%%%%%%%%%%%%%%%%%%%%%%%%%%%%%%%%%%%%%%%%%%%%%%%%%%
\section{Introduction}
The $f(R)$ gravity theory is considered as a representative theory
of modified gravities.
 The $f(R)$
gravity~\cite{NO,sf,FT,Nojiri:2010wj} has  much attentions as a
strong candidate for explaining the current accelerating
universe~\cite{SN1,SN2}. When choosing the Hu-Sawicki model~\cite{Hu:2007nk}, the theory could give
rise to the late time cosmic acceleration  without violating the gravity tests in the solar system and without affecting high redshift physics.
Very recently, the observational constraint on this model were reported from weal lensing peak abundances~\cite{Liu:2016xes}.
 Particularly, $f(R)=R+R^2/(6M^2)$
gravity~\cite{Starobinsky:1980te,Whitt:1984pd,Ferrara:2014ima} has
shown  a strong evidence for inflation to support  recent Planck
data~\cite{Ade:2015lrj}.  An important feature of this model
indicates  that the inflationary dynamics were driven by the purely
gravitational interaction  $R^2$ and the scale of inflation is
linked to the mass parameter $M^2$.
 This theory could thus  provide a unified picture of both
inflation in the early universe and the accelerated expansion at later times.
In addition, black hole~\cite{Multamaki:2006zb,delaCruzDombriz:2009et,Moon:2011hq}  and traversal wormhole solutions~\cite{Lobo:2009ip,Bronnikov:2010tt} have been found within $f(R)$ gravity in recent years.
The recent detection of gravitational waves by the LIGO Collaboration~\cite{Abbott:2016blz} is surely a milestone in gravitational waves research and opens new perspectives in the study of Einstein gravity (general relativity) and astrophysics.
Hence, it is meaningful to explore gravitational waves in the modified theory of gravity, especially in $f(R)$ gravity.
The observation of the polarization modes of gravitational waves will be a crucial tool to obtain valuable information about the black holes and the physics of the early universe.

It is well-known that the Einstein gravity  with
two polarization  degrees of freedom (DOF) is distinguished  from the metric $f(R)$
gravity with three DOF~\cite{BS}. Importantly, it is worth noting
that the Einstein equation derived from  $f(R)$ gravity contains
fourth-order derivative terms. A simple way to avoid a difficulty
dealing with the fourth-order equation is to transform the  $f(R)$
gravity into a scalar-tensor theory which is  surely a second-order
theory.

Very recently, it was reported that a polynomial $f(R)$ model could
provide two additional scalars of a massive longitudinal mode
(perturbed Ricci scalar: $R^{(1)}$) and a massless transverse mode
(breathing mode: $\hat{h}_b$), in addition to the two TT tensor
modes ($\hat{h}^+,\hat{h}^\times$)~\cite{Rizwana:2016qdq}. A
breathing mode seems to be overlooked in the literature because of
the assumption that the application of the Lorentz gauge  implies
the TT wave equation.  Also, it was insisted that
four DOF found in~\cite{Rizwana:2016qdq}  is consistent with the result obtained from the
Newman-Penrose (NP) formalism.  However, the presence of a breathing
mode contradicts  to the well-known fact in the literature that the
$f(R)$ gravity involves three DOF of a massive longitudinal mode and
two spin-2 modes. Hereafter, we wish to call this as the issue of
DOF in $f(R)$ theories.

In this work, we wish to point out that the $f(R)$ gravity still
involves three DOF by investigating  the fourth-order  equation
composed of  a second-order tensor and   a fourth-order scalar.

It seems that there is no breathing mode because  the perturbed
Ricci scalar $R^{(1)}$ is related closely to the trace `$h$' of
perturbed metric tensor. Hence, the allocation of the Ricci scalar
as a newly scalar represents   the the trace of metric tensor. Also,
we wish to   remind the reader that the Ricci scalar equation is not
an independent equation and is not separated from the perturbed
Einstein equation because it comes out just from taking the trace of
the latter equation.  This implies that the Ricci scalar is an
emergent mode from $h$. Furthermore, it is instructive to note that
in the TT gauge, there is  a close connection between the metric
perturbation and the linearized Riemann tensor, implying that
$\delta R_{itjt}=-\ddot{h}^{\rm TT}_{ij}/2$~\cite{Flanagan:2005yc}.
This gauge is very convenient because it fixes all local gauge
freedoms. But, it might be unclear  that there exists a close
relation between the metric perturbation and the NP formalism unless
one chooses the TT gauge.  One could not naively choose the TT gauge
in the perturbed $f(R)$ gravity because of $h\not=0$, whereas the
Lorentz gauge is easily implemented to eliminate the gauge DOF.
However, one might choose the TT gauge to obtain a massless spin-2
in the perturbed $f(R)$ gravity when one introduces a newly metric
perturbation $\tilde{h}_{\mu\nu}$.

The organization of our work is as follows. In the section 2, we briefly describe the  $f(R)$ gravity and its scalar-tensor theory
and derive two sets of perturbed equations around the Minkowski background in the section 3.
Sec.4 is focused on obtaining the number of  propagating DOF when one chooses the Lorentz gauge.
Finally, we will discuss our result which shows that there is no breathing mode in the section 5.

\section{$f(R)$ gravity and its scalar-tensor theory}
Instead of a polynomial model of  $f(R)=R+\alpha R^2+\beta
R^3+\cdots$~\cite{Myung:2011we,Rizwana:2016qdq}, we start   with  a
specific $f(R)$ gravity (Starobinsky
model~\cite{Starobinsky:1980te})
\begin{eqnarray}
S_{\rm f}=\frac{M^2_{\rm P}}{2}\int d^4 x\sqrt{-g} f(R),~~
f(R)=R+\frac{R^2}{6M^2},\label{Action}
\end{eqnarray}
where the $R^2$-term was originally motivated by one-loop correction
to Einstein gravity.  Here the mass parameter $M^2$ is chosen to be
a positive value, which is consistent with the stability condition
of $f''(0)>0$~\cite{sf}. This model of  $f(R)$ gravity is enough to
find the propagating DOF around the Minkowski background.  The
Einstein equation takes the form
\begin{eqnarray} \label{equa1}
R_{\mu\nu} f'(R)-\frac{1}{2}g_{\mu\nu}f(R)+
\Big(g_{\mu\nu}\nabla^2-\nabla_{\mu}\nabla_{\nu}\Big)f'(R)=0,
\end{eqnarray}
where the prime (${}^{\prime}$) denotes the differentiation with
respect to its argument.

 On the other hand, one might represent
(\ref{Action})  as a scalar-tensor theory  by introducing an
auxiliary field $\psi$~\cite{Whitt:1984pd}
\begin{equation} \label{a-Act}
S_{\rm A}=\int d^4x\sqrt{-g^J}\Big[\frac{M^2_{\rm
P}}{2}R+\frac{M_{\rm P}}{M}R\psi-3\psi^2\Big],
\end{equation}
where the superscript $J$ means the Jordan frame. Varying $S_{\rm
A}$ with respect to $\psi$ provides
\begin{equation} \label{psi-R}
\psi=\frac{M_{\rm P}}{6M} R
\end{equation}
which means that the Ricci scalar is treated as an independent
scalar degree of freedom. Plugging (\ref{psi-R})  into $S_{\rm A}$
again leads to the original $f(R)$ gravity (\ref{Action}) exactly.

Making use of  the conformal transformation and redefining the
scalar field ($\psi\to \phi$)
\begin{equation} \label{conf-trans}
g^J_{\mu\nu}\to \frac{1}{1+\frac{2\psi}{MM_{\rm P}}}g^E_{\mu\nu} \to
e^{-\sqrt{\frac{2}{3}}\frac{\phi}{M_{\rm P}}} g^E_{\mu\nu},
\end{equation}
one  arrives at  the Starobinsky model in the Einstein
frame~\cite{Ferrara:2014ima}
\begin{equation} \label{st-Act}
S_{\rm S}=\int d^4x\sqrt{-g^E}\Big[\frac{M^2_{\rm
P}}{2}R-\frac{1}{2} \partial_\mu \phi\partial^\mu \phi-V_{\rm
S}(\phi)\Big]
\end{equation}
with the Starobinsky potential
\begin{equation}
V_{\rm S}(\phi)=\frac{3M^2_{\rm
P}M^2}{4}\Big[1-e^{-\sqrt{\frac{2}{3}}\frac{\phi}{M_{\rm
P}}}\Big]^2.
\end{equation}
At this stage, we note that the conformal transformation
(\ref{conf-trans}) is a purely classified transformation of
coordinates and results in one frame are classically  equivalent to
the ones obtained other frame. Hence, it is plausible that the
number of DOF in the scalar-tensor theory (\ref{st-Act}) is three
because of two TT tensor modes and one scalar mode. From
(\ref{st-Act}), the Einstein and scalar equations are derived as
\begin{eqnarray}
\label{s-ein}&&G_{\mu\nu}=\frac{1}{M^2_{\rm
P}}T^{\phi}_{\mu\nu},~~T^\phi_{\mu\nu}=\partial_\mu\phi\partial_\nu\phi-\frac{1}{2}g_{\mu\nu}\Big[(\partial\phi)^2+V_{\rm
S}\Big],
\\
\label{s-s}&&\nabla^2\phi-V'_{\rm S}=0,~~V'_{\rm
S}=\sqrt{\frac{3}{2}} M_{\rm
P}M^2e^{-\sqrt{\frac{2}{3}}\frac{\phi}{M_{\rm
P}}}\Big[1-e^{-\sqrt{\frac{2}{3}}\frac{\phi}{M_{\rm P}}}\Big].
\end{eqnarray}
The above describes a process of [$R^2\to R\psi-3\psi^2\to
-(\partial \phi)^2-V$] briefly.

\section{Two sets of perturbed equations}
 We introduce  the metric perturbation around the Minkowski background to
find out the propagating DOF
\begin{eqnarray} \label{m-p}
g_{\mu\nu}=\eta_{\mu\nu}+h_{\mu\nu}.
\end{eqnarray}
 The Taylor expansions around $R=0$ are  employed  to
define the linearized Ricci scalar $\delta R(h)$
as~\cite{Myung:2011ih}
\begin{eqnarray}
f(R)&=& f(0)+f'(0)\delta R(h)+\cdots, \\
f'(R)&=& f'(0)+f''(0)\delta R(h)+\cdots
\end{eqnarray}
with  $f(0)=0$,  $f'(0)=1$, and $f''(0)=1/3M^2$. We note that
$\delta R(h)$ will be used here instead of $R^{(1)}$
in~\cite{Rizwana:2016qdq}.
 The perturbed
(linearized) equation to (\ref{equa1}) is given by the fourth-order
coupled equation
\begin{eqnarray}
\delta R_{\mu\nu}(h)+\frac{1}{3M^2}
\Big[\eta_{\mu\nu}\Big(-\frac{3M^2}{2}+\square\Big)-\partial_{\mu}\partial_{\nu}
\Big]\delta R(h)=0,~~\square=\partial^2, \label{leq}
\end{eqnarray}
where the linearized Ricci tensor and  scalar are given by
\begin{eqnarray}
&&\delta
R_{\mu\nu}(h)=\frac{1}{2}\Big[\partial^{\rho}\partial_{\mu}h_{\nu\rho}+
\partial^{\rho}\partial_{\nu}h_{\mu\rho}-\square
h_{\mu\nu}-\partial_{\mu}
\partial_{\nu}h\Big],\label{lRmunu}\\
&&\delta R(h)=\partial^{\rho}\partial^{\sigma}h_{\rho\sigma}
-\square h.\label{lR}
\end{eqnarray}
When using (\ref{lRmunu}) and (\ref{lR}), the linearized equation
(\ref{leq}) becomes  a second (fourth)-order differential equation
with respect to  $h_{\mu\nu}(h)$. Obviously, it is not a tractable
equation. Furthermore, its trace equation leads to the linearized
Ricci scalar equation
\begin{equation} \label{th-eq2}
 (\square-M^2)\delta R=0.
 \end{equation}
Introducing the linearized Einstein tensor $\delta G_{\mu\nu}=\delta
R_{\mu\nu}-\eta_{\mu\nu} \delta R/2$, Eq.(\ref{leq}) takes a compact
form
\begin{eqnarray}
\delta G_{\mu\nu}(h)+\frac{1}{3M^2}
\Big[\eta_{\mu\nu}\square-\partial_{\mu}\partial_{\nu} \Big]\delta
R(h)=0. \label{sleq}
\end{eqnarray}
We note that the Bianchi identity is satisfied when acting
$\partial^\mu$ on (\ref{sleq}).

 On the other hand, two linearized equations from
(\ref{s-ein}) and (\ref{s-s}) together with $\phi=0+\varphi$ take
the simple forms with $\delta T^{\phi}_{\mu\nu}=0$ and $\delta R=0$
\begin{eqnarray}
\label{phi-eq1}&&\delta R_{\mu\nu}(h)=0, \\
\label{phi-eq2}&& (\square-M^2)\varphi=0.
\end{eqnarray}
We note that  Eqs.(\ref{th-eq2}) and (\ref{phi-eq2}) are the same
when replacing $\delta R$ by $\varphi$, but the fourth-order coupled
equation (\ref{leq}) is quite different from the linearized Einstein
equation (\ref{phi-eq1}). This indicates that  (\ref{leq}) can be
reduced to two decoupled second-order equations (\ref{phi-eq1}) and
(\ref{phi-eq2}) if one employs the conformal transformation and
redefinition of scalar appropriately after choosing (\ref{a-Act}),
leading to a canonical scalar action with the Starobinsky  potential
in the Einstein frame. In the scalar-tensor theory approach, one
assigns the perturbed Ricci scalar to  an independent scalar
$\varphi$. Instead, it does not have the trace of metric
perturbation $h$.

\section{Propagating DOF with the Lorentz gauge}
In order to take into account the propagating DOF in $f(R)$ gravity,
it is convenient to separate the metric tensor $h_{\mu\nu}$ into the
traceless part $h^{T}_{\mu\nu}$ and the trace part $h$ as
\begin{equation} \label{h-ttpart}
h_{\mu\nu}=h^{\rm T}_{\mu\nu}+\frac{h}{4}\eta_{\mu\nu}
\end{equation}
with $h^{{\rm T} \mu}_{~~~\mu}\equiv h^{\rm T}=0$. This splitting is
meaningful because the issue of DOF in $f(R)$ theories is related
to the presence of $h$.

First of all, let us choose the Lorentz (harmonic) gauge to
eliminate the gauge DOF
\begin{equation} \label{lorentz-g}
\partial_\mu h^{\mu\nu}=\frac{1}{2}\partial^\nu h \to \partial_\mu h^{{\rm T}\mu\nu}=\frac{1}{4}\partial^\nu
h. \end{equation} Here, we note that the transverse condition of
$\partial_\mu h^{{\rm T}\mu\nu}=0$ cannot be achieved in $f(R)$
gravity because of  $h\not=0$.
 Then, the linearized Ricci tensor and scalar
are given by
\begin{equation} \label{Ricci-peq}
\delta R(h)_{\mu\nu}\equiv \delta R^{\rm
T}_{\mu\nu}+\frac{\eta_{\mu\nu}}{4}\delta R
=-\frac{1}{2}\Big[\square h^{\rm
T}_{\mu\nu}+\frac{\eta_{\mu\nu}}{4}\square h\Big],~~\delta
R(h)=-\frac{1}{2}\square h,
\end{equation}
where the last equation indicates that the linearized Ricci scalar
exists iff $h\not=0$ under the Lorentz gauge. This implies that
$\delta R$ cannot be defined without $h$. That is,  if $h=0$,
$\delta R=0$.

 Now, the fourth-order
equation (\ref{leq}) leads to
\begin{equation} \label{4tho-eq}
\square
h_{\mu\nu}+\frac{1}{3M^2}\Big[\eta_{\mu\nu}\Big(-\frac{3M^2}{2}+\square\Big)-\partial_\mu\partial_\nu\Big]\square
h=0.\end{equation} The other form of (\ref{4tho-eq}) takes the form
\begin{equation} \label{4th-eq}
\square h^{\rm
T}_{\mu\nu}+\frac{1}{3M^2}\Big[\eta_{\mu\nu}\Big(-\frac{3M^2}{4}+\square\Big)-\partial_\mu\partial_\nu\Big]\square
h=0.\end{equation}

If $\square h\not=0$, its  trace equation takes the form
\begin{equation} \label{h-4th}
(\square-M^2)\square h=0 \end{equation} which is actually the same
equation as in (\ref{th-eq2}). Here, Eq.(\ref{h-4th}) implies
\begin{equation} \label{h-2nd}
(\square -M^2)h=0
\end{equation}
because $h$  could  represent a massive (scalar) graviton mode in
$f(R)$ gravity. The other case of $\square h=0$ is not allowed since
if $\square h=0$, one could not derive the trace equation
(\ref{h-4th}) itself. Importantly, this issue should be carefully
treated because the massless mode satisfying $\square h = 0$ may
correspond to the breathing mode, which is the main subject of this
work.  In general, it seems that the solution of Eq.(\ref{h-4th}) is
given by the sum of the massive mode and  massless mode which are
independent with each other. However, the massless mode which is a
solution to $\square h=0$ does not exist in $f(R)$ gravity.
 We stress that
$h$ plays the role of a propagating massive mode instead of $\delta
R$. Here is the additional  reason  to understand why the massless mode (breathing
mode) cannot survive in $f(R)$ gravity. If one requires $\square
h=0[\delta R=0$, via (\ref{Ricci-peq})], Eq.(\ref{4th-eq}) reduces
to $\square h^{\rm T}_{\mu\nu}=0$, which is just the tensor equation
in Einstein gravity when choosing the Lorentz gauge. It is worth
noting that the last fourth-order term of (\ref{4th-eq}) indicates a
feature of the perturbed $f(R)$ gravity clearly. If one chooses
$\square h=0$, this term disappears.  Therefore, we clarify that the
massless scalar mode does not exist.

 Acting $\partial^\mu$ on (\ref{4th-eq}) leads
to (\ref{lorentz-g}), which implies that the Lorentz-gauge condition
is satisfied in the perturbed equation level. Plugging (\ref{h-4th})
into (\ref{4tho-eq}), we have
\begin{equation} \label{4tho-seq}
\square
h_{\mu\nu}-\frac{1}{3M^2}\square\Big[\frac{M^2}{2}\eta_{\mu\nu}+\partial_\mu\partial_\nu\Big]
h=0.\end{equation}
 Substituting
(\ref{h-4th}) into (\ref{4th-eq}) implies a fourth-order equation
\begin{equation} \label{4th-seq}
\square h^{\rm
T}_{\mu\nu}+\frac{1}{3M^2}\square\Big[\frac{M^2}{4}\eta_{\mu\nu}-\partial_\mu\partial_\nu\Big]
h=0\end{equation} which shows clearly  that the traceless metric
perturbation  $h^{\rm T}_{\mu\nu}$ is closely coupled to the trace
of metric  perturbation $h$. It is clear  that the trace mode $h$
cannot be decoupled from the traceless tensor mode $h^{\rm
T}_{\mu\nu}$. This is the origin of difficulty met when taking into
account  DOF arisen from the $f(R)$ gravity. Interestingly,
Eq.(\ref{4th-seq}) can be  transformed  to the Ricci tensor-Ricci
scalar equation
\begin{equation}
 \label{4th-RReq}
\delta R^{\rm
T}_{\mu\nu}+\frac{1}{3M^2}\Big[\frac{M^2}{4}\eta_{\mu\nu}-\partial_\mu\partial_\nu\Big]\delta
R=0,\end{equation} which indicates that the traceless Ricci tensor
is coupled to the Ricci scalar.

At this stage, we observe that Eq.(\ref{4th-seq}) may become a
massless propagating tensor equation for $\tilde{h}_{\mu\nu}$ as
\begin{equation} \label{th-eq}\tilde{h}_{\mu\nu}\equiv h^{\rm
T}_{\mu\nu}+\frac{1}{3}\Big(\frac{\eta_{\mu\nu}}{4}-\frac{\partial_\mu\partial_\nu}{M^2}\Big)h,~~\square \tilde{h}_{\mu\nu}=0,
\end{equation}
which suggests  a way of defining a massless spin-2 in $f(R)$
gravity. That is, plugging $h^{\rm T}_{\mu\nu}$ in (\ref{th-eq})
into (\ref{4th-seq}) leads to $\square \tilde{h}_{\mu\nu}=0$.  We
find  from (\ref{h-ttpart}) that $h^{\rm T}_{\mu\nu}$ differs from
$h_{\mu\nu}$ by $h$. Especially, the splitting of $h$ in
$\tilde{h}_{\mu\nu}$ is nontrivial, which reflects a feature of
$f(R)$ gravity. This is compared to that of $h_{\mu\nu}$ in Einstein
gravity. For this purpose, we may express $\tilde{h}_{\mu\nu}$ as
\begin{eqnarray}
\tilde{h}_{\mu\nu}&=&h_{\mu\nu}-\frac{1}{3}\Big(\frac{\eta_{\mu\nu}}{2}+\frac{\partial_\mu\partial_\nu}{M^2}\Big)h \label{tilde-h}\\
&=&\bar{h}_{\mu\nu}-\frac{1}{3}\Big(\eta_{\mu\nu}-\frac{\partial_\mu\partial_\nu}{M^2}\Big)\bar{h},
\label{bar-h}
\end{eqnarray}
where $\bar{h}_{\mu\nu}=h_{\mu\nu}-\eta_{\mu\nu}h/2$ is the
trace-reversed metric perturbation
$(\bar{h}=-h)$~\cite{Flanagan:2005yc}. Here, the Lorentz gauge is
given by $\partial^\mu\bar{h}_{\mu\nu}=0$.  We may rewrite
(\ref{4tho-eq}) in term of $\bar{h}_{\mu\nu}$
\begin{equation}
 \label{4th-bh}
\square
\bar{h}_{\mu\nu}-\frac{1}{3M^2}\Big[\eta_{\mu\nu}\square-\partial_\mu\partial_\nu\Big]
\square \bar{h}=0,\end{equation} which is surely the same equation
found in Ref.\cite{BS}.  Substituting $\bar{h}_{\mu\nu}$  defined in
(\ref{bar-h}) into  that in (\ref{4th-bh}) arrives at $\square
\tilde{h}_{\mu\nu}=0$ which is the same equation as in
(\ref{th-eq}).

 Using the trace equation (\ref{h-2nd})
and the Lorentz gauge (\ref{lorentz-g}), we may impose  the TT
condition for $\tilde{h}_{\mu\nu}$
\begin{equation} \label{l-tt}
\partial_\mu \tilde{h}^{\mu\nu}=0,~~\tilde{h}^{\mu}~_{\mu}=0,
\end{equation}
which indicates that $\tilde{h}_{\mu\nu}$ is a newly tensor mode
defined in $f(R)$ gravity.  This may imply that $f(R)$ gravity
accommodates three DOF of two from $\tilde{h}_{\mu\nu}$ and one from
$h$. We note that if $h=0$, $\tilde{h}_{\mu\nu}$ reduces to $h^{\rm
T}_{\mu\nu}$ and to $ h^{\rm TT}_{\mu\nu}$ finally, leading to
Einstein gravity.

Considering a  gravitational wave that propagates in the $z$
direction, Eq.(\ref{th-eq}) together with (\ref{l-tt})  exhibits the
fact that gravitational waves have two polarization components.
Explicitly, Eq.(\ref{l-tt}) implies
\begin{equation}
\tilde{h}_{tt}=\tilde{h}_{ti}=0,~~\tilde{h}^i~_i=0,~~\partial^i\tilde{h}_{ij}=0,
\end{equation}
where the last two expressions correspond to  the TT gauge.
 $\tilde{h}_{ij}^{\rm TT}=\tilde{h}_{ij}^{\rm TT}(t-z)$ is
a valid solution to the TT wave equation $\square
\tilde{h}_{ij}^{\rm TT}=0$. The TT gauge condition of $\partial_z
\tilde{h}_{zj}^{\rm TT}=0$ implies $\tilde{h}_{zj}^{\rm
TT}(t-z)$=constant and however, this component should be zero to
satisfy  a condition of the asymptotic flatness:
$\tilde{h}_{\mu\nu}\to 0$ as $z\to \infty$. The remaining non-zero
components of $\tilde{h}_{ij}^{\rm TT}$ are given by
$\tilde{h}_{xx}^{\rm TT},~\tilde{h}_{xy}^{\rm
TT},~\tilde{h}_{yx}^{\rm TT}$ and $\tilde{h}_{yy}^{\rm TT}$.
Requiring the symmetry and traceless condition leads to the  two
independent components
\begin{equation}
\tilde{h}_{xx}^{\rm TT}=-\tilde{h}_{yy}^{\rm TT}\equiv
\tilde{h}^+(t-z),~~\tilde{h}_{xy}^{\rm TT}=\tilde{h}_{yx}^{\rm
TT}\equiv\tilde{ h}^\times(t-z).
\end{equation}
Now, considering (\ref{h-2nd}), one find the trace
solution~\cite{Capozziello:2008rq}
\begin{equation}
h=h^0e^{ik_\mu x^\mu} \to h(t-v_G z)
\end{equation}
where $v_G=k/\omega=\sqrt{\omega^2-M^2}/\omega<1(\omega^2=M^2+k^2)$
is the group velocity of a massive scalar graviton.

Finally, we obtain  $h_{\mu\nu}$ as the solution to (\ref{4tho-seq})
with (\ref{tilde-h})
\begin{equation} \label{h-solution}
h_{\mu\nu}(t,z)=\tilde{h}^+(t-z)e^{(+)}_{\mu\nu}+\tilde{h}^\times(t-z)e^{(\times)}_{\mu\nu}
+\frac{1}{3}\Big(\frac{\eta_{\mu\nu}}{2}-\frac{k_\mu
k_\nu}{M^2}\Big)h(t-v_Gz).
\end{equation}
The other solution $\bar{h}_{\mu\nu}$ as the solution to
(\ref{4th-bh}) with (\ref{bar-h}) takes the form
\begin{equation} \label{hb-solution}
\bar{h}_{\mu\nu}(t,z)=\tilde{h}^+(t-z)e^{(+)}_{\mu\nu}+\tilde{h}^\times(t-z)e^{(\times)}_{\mu\nu}
+\frac{1}{3}\Big(\eta_{\mu\nu}+\frac{k_\mu
k_\nu}{M^2}\Big)h(t-v_Gz),
\end{equation}
which is  the same solution found in Ref.\cite{BS}. This encodes
that three DOF of ($\tilde{h}^{+},\tilde{h}^{\times},h$) are found
from the $f(R)$ gravity.

In order to compare (\ref{h-solution}) and (\ref{hb-solution}) with
the Ricci scalar-like
solution~\cite{Capozziello:2008rq,Berry:2011pb,Rizwana:2016qdq}, we
write down its solution by replacing $\delta R=-\square h/2$ with
$h_f$
\begin{equation} \label{cap-solution}
h_{\mu\nu}^{\rm
R}(t,z)=A^+(t-z)e^{(+)}_{\mu\nu}+A^\times(t-z)e^{(\times)}_{\mu\nu}
+h_f(t-v_Gz)\eta_{\mu\nu}.
\end{equation}
Even though  ($A^{+},A^{\times},h_f$) are similar to
($\tilde{h}^{+},\tilde{h}^{\times},h$), the last term of solution
$h_{\mu\nu}^{\rm R}$ (\ref{cap-solution}) differs from that of
$h_{\mu\nu}$ (\ref{h-solution}).

 On the other side of the scalar-tensor theory, Eq.(\ref{phi-eq1})
together with $\delta R=0$ reduces to
\begin{equation}
\delta R^{\rm T}_{\mu\nu}=0  \to \square h^{\rm T}_{\mu\nu}=0.
\end{equation}
Also, as was shown in Eq.(\ref{phi-eq2}), the scalar mode $\varphi$
is decoupled completely from the $h^{\rm T}_{\mu\nu}$. Requiring the
transverse condition of $\partial_\mu h^{{\rm T}\mu\nu}=0$ (Lorentz
condition with $h=0$) leads to two DOF of $(h^+,h^\times)$, which
describe the general relativity.  Hence, it is obvious that the
scalar-tensor theory has three DOF (one scalar DOF+ two tensor
DOF).

\section{Discussions}

First of all, we note that three DOF of
($\tilde{h}^{+},\tilde{h}^{\times},h$) (\ref{h-solution})  are found
from analyzing the perturbed  $f(R)$ gravity. Here, we did not
introduce the Ricci scalar mode ($\delta R=R^{(1)}$) separately
because it is closely related to the trace of metric tensor $h$. We
have solved the fourth-order coupled equation (\ref{4tho-seq})
together with the trace equation (\ref{h-2nd}) directly.

We have found that there is no breathing mode in $f(R)$ gravity. The
four DOF including breathing mode have been obtained
in~\cite{Rizwana:2016qdq} by assuming that the traceless condition
of $h=0$ cannot be imposed  on the perturbed $f(R)$ gravity, after
counting the Ricci scalar mode.  The authors in~\cite{Rizwana:2016qdq} have discovered the breathing
mode ($\hat{h}_b$) from the condition of $h\not=0$ when the
background spacetime is not Minkowski. The approach used in
\cite{Rizwana:2016qdq} was based on the observation that $R^{(1)}$
is considered  as a different mode from $h$
initially~\cite{Berry:2011pb}. This might lead to overcounting of
DOF. However, noting  an expression of $\delta R=-\frac{\square
h}{2}$ (\ref{Ricci-peq}) in the Lorentz gauge implies that $\delta
R$ is closely connected to  $h$.

One might attempt to argue from (\ref{h-4th}) that the massless mode
satisfying $\square h = 0$ may correspond to the breathing mode.  In
general, it seems that the solution to the fourth-order equation
(\ref{h-4th}) is given by the sum of the massive mode and massless
mode which are independent with each other.  However, the massless
mode which is a solution to $\square h=0(h\not=0)$ does not exist in
$f(R)$ gravity since  $\square h=0$ [via (\ref{Ricci-peq})] means
$\delta R=0$ in the Lorentz gauge.

Consequently, we have clarified  the the issue of DOF in $f(R)$
theories. The number of polarization modes of gravitational waves in
$f(R)$ gravity is still `three' in  Minkowski spacetime, which is
consistent with the results in the literature (especially
for~\cite{BS}). Also, we would like to mention that the  DOF
counting of $f(R)$ theories should be  independent of propagating
spacetime.

\section*{Acknowledgement}
This work was supported by the 2016 Inje University research grant.

\end{document}